# Inspecting plastic deformation of Pd by means of fractal geometry

Ali Eftekhari *

*Electrochemical Research Center, P.O. Box 19395-5139, Tehran, Iran*

**Abstract**

The influence of phase transformation-induced plastic deformation in Pd|H system on the electrode surface was investigated. Since the Pd surface is subject of severe plastic deformation during this process, the structure and roughness of the electrode surface significantly changes. Quantitative analysis of the electrode surfaces for comparative study of such changes is a valuable tool to inspect the plastic deformation induced. Fractal dimension can be used as a quantitative measure for this purpose. Since inappropriate methods may lead to significant errors, an appropriate approach was proposed for the determination of fractal dimensions in such systems. It was demonstrated that the surface roughness generated is mainly due to the plastic deformation induced, not the other side processes. The methodology is of general interest for the investigation of plastic deformations. The methodology is of general interest for the investigation of plastic deformations, and also other solid-state structural changes.

* Corresponding author. Tel.: +98-21-204-2549; fax: +98-21-205-7621.

*E-mail address:* eftekhari@elchem.org.

# 1. Introduction

Adsorption of hydrogen on palladium is different of that on other metals, since it is accompanied by the Penetration reaction. This means that the adsorbed hydrogen atoms penetrate the Pd surface layer and moves into an internal bulk site [1]. Thus, Pd|H system has achieved a particular attention, because palladium is able to store a huge amount of hydrogen [2]. Hydrogen injection into and subsequent extraction from a Pd electrode is accompanied by phase transformations between $\alpha$-PdH and $\beta$-PdH [3,4]. Such phase transformation results in severe plastic deformation of the Pd electrode. As a result, the structure and roughness of the Pd surface significantly changes. Investigating such surface changes is of interest from both fundamental and applied points of view. It is of interest from solid-state physics point of view to understand the plastic deformation induced [5], and also for applied purposes, since such surface changes significantly affect electrochemical behavior of Pd electrode [6].

Pyun and his co-workers [7] have recently used fractal geometry for the investigation of such plastic deformation induced to Pd electrode. This is indeed a brilliant approach for this purpose. However, this needs some clarification of the methodology and modification of the approach to generalize it for vast applications. Upon this action, this method can be generally used to monitor solid-state structural changes, not only plastic deformation.

In a previous paper [8], it has been shown that electrochemical methods can be used for fractal analysis of almost all classes of materials in solid-state physics. In the present paper, it is aimed to show the powerfulness of electrochemical methods to inspect solid-state structural changes of solids via fractal analysis. To this aim, plastic deformation of Pd was chosen as a typical example, since it is well-known in both electrochemistry and solid-state physics. For sake of clarity, this typical case is discussed throughout the present study, and the results can be simply extended to other similar cases in solid-state physics.

## 2. Experimental

For the sake of similarity of the experimental results, the experiments were performed in accordance with the procedure reported in [7]. Briefly, different amounts of hydrogen were injected into Pd electrodes by applying a constant current of 13.5 mA cm$^{-2}$ for various times under galvanostatic condition in a solution of 0.1 M NaOH. Then, all Pd electrodes were discharged to extract the whole hydrogen injected. This process was carried out potentiostatically by jumping the electrode potential from 0.5 to 0.9 V vs. SCE. Similar to [7], four different Pd electrodes were prepared by injection of various amounts of hydrogen during the preliminary charging process, as the galvanostatic current applied for different times: 0, 0.9, 1.8 and 2.7 ks (10$^3$ s). Thus, different Pd electrodes approximately with the pre-charged states of PdH$_0$, PdH$_{0.23}$, PdH$_{0.46}$ and PdH$_{0.6}$ were obtained. Note that these are just symbols employed to distinguish them, and indeed all of them were discharged to the state of Pd. The PdH$_0$ is a Pd electrode without any plastic deformation, and PdH$_{0.6}$ is considered as a Pd electrode with severe plastic deformation.

For the fractal analysis of the Pd electrodes, they were covered by Au masks. The ideal masking is the deposition of monolayer of Au, but a little more charge was used to assure about the completeness of masking. The thickness of the gold mask was about 2 nm. Therefore, the Au-masked Pd electrode can be considered as conventional electrodes, and their fractal dimensions were determined using sufficiently fast redox of Fe(CN)$_6^{3-}$/Fe(CN)$_6^{4-}$ at gold surface. The electrolyte was an aqueous solution of 3 M NaCl and 15 mM K$_4$Fe(CN)$_6$ and 15 mM K$_3$(CN)$_6$. The experimental details can be found in the previous papers [9-11].

It should be taken into account that process of plastic deformation is not important in this study. Indeed, plastic deformations were generated electrochemically, since it is a simple method for this purpose; however, any other processes can be used to generate plastic deformations. In other words, terminology of Pd electrode does not restrict the method to electrochemical systems, and any Ph specimen can be analyzed to inspect the plastic deformation.

## 3. Results and discussion

*3.1. The problems of conventional method for fractal analysis*

It is known that the reason for the appearance of such plastic deformations is phase transformation of $\alpha$-PdH to $\beta$-PdH, and thus this issue will be disregarded here. In addition, it has been reported that the fraction of $\beta$-PdH is responsible for the strength of the plastic deformation, and more injected hydrogen in the charging process results in higher fraction of $\beta$-PdH. Thus, the four electrodes viz. $PdH_0$, $PdH_{0.23}$, $PdH_{0.46}$ and $PdH_{0.6}$ can be considered as samples with different strengths of plastic deformation induced. In other words, regardless of the electrochemical procedure inducing such plastic deformations, these samples will be used throughout this study to investigate the degree (strength) of plastic deformation.

Pyun and his co-workers [7] have used conventional fractal analysis of electrode surfaces based on electrochemical methods to measure the fractal dimensions of the Pd electrodes (with different strengths of plastic deformation) varying from 1.97 to 2.09. There are two problems for the results reported in [7]. *(i)* The values of the fractal dimensions estimated are significantly lower than the real ones, as it is expected to be significantly higher than 2 (since they are rough surfaces), e.g., from 2.2 to 2.5. *(ii)* The difference of the fractal dimension of the Pd electrodes with and without plastic deformation is significantly low. Some much the worse, some values suggest dimensions between one and two and some between two and three, which are not applicable for comparative study due to the lack of geometrical meaning. It is known that Pd surface has significant roughness in microscopic scale, which this roughness strongly increases when the Pd is subject of electrochemical treatment. Thus, it is believed that the fractal dimension of the first Pd electrode should be higher than 2, e.g. 2.2. On the other hand, the $PdH_0$ electrode was subject of no plastic deformation, and the $PdH_{0.6}$ subject of severe plastic deformation. Only a 0.13 increase in the fractal dimensions as a result of such severe plastic deformation is not acceptable.

The authors of [7] have claimed that the reason for such low values of $D_f$ could be attributed to the formation of PdO, which is not electrochemically active and does not participate in the electrochemical reaction to monitor the fractal dimension. This is an acceptable assumption, since the process of "diffusion towards electrode surface" just senses a part of the electrode surface which is electrochemically active. This means that a part of the electrode surface, which is electrochemically inactive, is missed. Unfortunately, this is an important part of the electrode surface, which is missed. If only a part of the Pd surface was subject of oxidation to form PdO, this part should probably be the deformed part which is sensitive to participate in an electrochemical reaction.

There is another problem regarding the reliability of conventional electrochemical methods for the fractal analysis of the system under investigation. Since, the fractal dimension is calculated from the electrical signals received from the electrode, it is very important to avoid side-processes to obtain accurate data. First, plastic deformation due to phase transformation is slightly different of simple roughening of electrode surfaces. This process can be accompanied by the formation of interfacial defects, which have significant effect on electrochemical reactions. On the other hand, in conventional method employed in [7], the diffusing species for the determination of the fractal dimension of the electrode surface is hexacyanoferrate ion. Whereas, it is known that hexacyanoferrate ions can react with metallic Pd to form palladium hexacyanoferrate, even electrolessly [12,13].

*3.2. Fractal analysis by gold-masking approach*

Now, it is appropriate to propose an alternative approach instead of the conventional method of the $Fe(CN)_6^{3-}/Fe(CN)_6^{4-}$ redox couple at the original electrode surface. In a series of papers [9-11,14], we have shown that gold masking of the electrode surface and then estimating the fractal dimension of the Au-covered electrode is an efficient and reliable approach for the determination of the fractal dimensions of electrode surfaces. This approach was successfully employed to monitor fractal structure of different surfaces for comparative studies in different systems: electroactive materials (to investigate cycling and aging effects) [9], alloys (to inspect corrosion) [10], non-conducting materials such as teeth (to monitor dental decay) [1`], and even liquid | liquid interfaces [14]. For an original Pd electrode, all parts of the electrode surface do not participate in the electrochemical reaction; thus, fractal analysis of the electrode surface is just restricted to its electroactive part. By gold-masking approach, the electrode surface becomes uniformly electroactive in all parts.

One may think about two possible problems associated with gold-masking approach: *(i)* covering some fractalities of the original surface, and *(ii)* formation of Au own roughness. Fortunately, these problems are not important. It is well known that fractal objects have not fractality in all length scales [15]. Gold masking may just cover roughness of the original surface in very very small scale in comparison with the size of Au atom. On the other hand, ideal monolayer gold deposition may be accompanied by the formation of Au own roughness in the same scale (i.e., just atomic inhomogeneities). This is the reason for recommending monolayer gold-masking; though, gold masking with only a few layers thickness has not significant deviation from this assumption. In any case, regardless of the surface fractality scale, electrochemical methods based on "diffusion toward electrode surfaces" are useful for monitoring of fractality at length scales larger than one micron, and at the best conditions, just able to sense the distances higher than 100 nm. Thus, both the original surface irregularities covered during gold masking and the atomic inhomogeneities of the Au mask are not sensed by the diffusing species, and have no significant effect on the fractality monitored.

Since the diffusing species can sense the electrode surface structure, electrochemical methods can be used for the determinations of the electrode surfaces. Two reliable methods are scan rate dependency of the peak current and time dependency of the current in cyclic voltammetry and chronoamperometry, respectively. It is known that the diffusion-limited current in chronoamperometric measurements is proportional to the fractal parameter $\alpha$ as: $I(t) \propto t^{-\alpha}$ [16,17]. Similarly, the peak current recorded in cyclic voltammogram is dependent on the scan rate as: $I_p \propto v^{\alpha}$ [18,19]. The fractal parameter can be simply transformed to the fractal dimension by the equation $\alpha = (D_f - 1)/2$ [17]. Thus, the fractal dimension can be calculated by plotting the peak current of different cyclic voltammograms as a function of scan rate, or the diffusion-limited current versus time in logarithmic scales. In other words, slopes of such logarithmic plots are equal to the fractal parameter $\alpha$. Typical curves for the roughest Pd electrode, which was subject of severe plastic deformation due to phase transformation in $PdH_{0.6}$, are illustrated in Fig. 1.

By similar procedures, the fractal dimensions of different Pd electrodes were estimated (Table 1). The results indicate that the fractal dimension is proportional to the degree of plastic deformation induced, since the value of fractal dimension is higher for the Pd electrodes subjected to stronger plastic deformations. Therefore, the fractal dimension of the electrode surface can be used as a quantitative measure of the plastic deformation induced, but only in a comparative study. It is obvious that the standard deviation of the values calculated is lesser than that reported in [7]. This is due to the fact that the redox system at gold surface has a more ideal behavior in comparison with Pd surface. This is indeed another evidence for efficiency of the approach proposed.

The reliability of the results is not questionable, since the Pd electrodes act as conventional Au electrodes due to the Au uniform masks on them. Analysis of the results shows possibility of the approach employed to inspect the plastic deformation induced to the Pd electrodes. The fractal dimension estimated for the Pd electrode subjected to only electrochemical treatment is an acceptable value for such metallic electrodes. The value reported for the Pd electrode subjected to severe plastic deformation due to strong phase transition suggests a strongly rough surface, which is indeed what expected. The results obviously indicate that the plastic deformation can lead to the formation of significantly rougher surfaces with about 0.3 higher fractal dimension (under the experimental conditions of the present study).

### 3.3. Examining the reliability of the method

According to the recent achievements, fractal analysis is an efficient tool in electrode to monitor the structure of electrode surfaces. Even, it has been recently reported that fractal analysis can be used for study of electrochemical processes, as every reaction has a specified fractal dimension [20]. In any case, it has been emphasized that fractal dimension is just a number and without physical significance it is useless. Thus, fractal analysis is generally applicable for comparative studies. In this direction, it is very important to assure that the experiments performed at different conditions are actually the same.

Although, reliability of neither the gold-masking approach nor the fractal analysis is doubtful, there is still an important open issue that whether the fractal analysis of the Pd surface reflects the plastic deformation induced. It is obvious that plastic deformation is accompanied by surface changes; however, the influence of other effects should also be investigated. Since, the system under investigation is accompanied by complicated electrochemical processes at the electrode surface, it is of importance to assure that the fractality is exclusively (or mainly) due to the plastic deformation induced as a result of phase transformation, not simple surface processes. For instance, the formation of metal oxide on the electrode surface, as addressed in [7], seems to be a source of roughness of the electrode surface.

Pyun and his co-workers [7] have used an excessively anodic potential to assure about complete extraction of hydrogen from Pd, since the presence of even small amount of hydrogen could participate in the redox couple employed for the fractal analysis, which might be a source of error. Fortunately, this obligation is not applicable for the approach proposed here, since the Pd electrode is not directly in contact with the $Fe(CN)_6^{3-}/Fe(CN)_6^{4-}$ redox couple. In fact, this is another advantage of the gold-masking approach. Therefore, the extraction process can be performed by applying lesser anodic potentials. Incompleteness of the extraction process leading to the existence of a tiny amount of hydrogen is not problematic, as the plastic deformation occurs at high content of hydrogen, and small amount of hydrogen is purely in $\alpha$-PdH phase without any phase transformation.

The results obtained from the experimental measurements (Table 2) indicate that the fractal dimension of the Pd electrode discharged at lesser anodic potentials is slightly lower, probably due to the non-existence of metal oxide film. However, this difference is not significant, indicating that even the existence of oxide films is not responsible for the generation of such rough surfaces.

On the other hand, different processes occurring in the course of hydrogen desorption may be responsible for the formation of such rough surfaces of discharged Pd electrodes. It is known that hydrogen desorption from Pd is accompanied by several processes [21]:

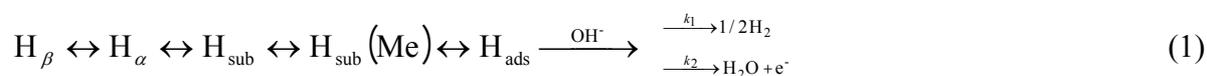

$$H_\beta \leftrightarrow H_\alpha \leftrightarrow H_{sub} \leftrightarrow H_{sub}(Me) \leftrightarrow H_{ads} \xrightarrow{OH^-} \begin{array}{c} \xrightarrow{k_1} 1/2H_2 \\ \xrightarrow{k_2} H_2O + e^- \end{array} \qquad (1)$$

The plastic deformation is due to the phase transformation occurring at the first process. The other processes are highly dependent on the experimental conditions employed for discharging. Since, the mechanism of hydrogen desorption is influenced by electrochemical changes induced by varying the potential scan rate, it is useful to compare the surface structures of the Pd electrodes discharged potentiodynamically at different potential scan rates. The results are reported in Table 3. This indicates significance of the first process.

### 3.4. Estimating the fractality scale

In addition to fractal dimension, inner and outer cutoffs the fractal surface can be estimated by means of the electrochemical methods. In fact, inner and outer cutoffs indicate the boundaries of fractality scale, since it is known from the concept of fractal geometry that fractal objects have not fractal structure in all length scales [15]. Thus, it is useful to estimate the scale of fractality for better understanding of the fractal structure of the surface. This is also of interest for comparative study of similar surface, such as the case under investigation.

As the diffusion layer acts as a yardstick length, the length scale relevant for the peak current can be estimated by calculating diffusion layer at the time $\tau$ when the current reaches its maximum value ($i_{peak}$). According to the Fick's first law, the diffusion-limited current is proportional to the magnitude of the concentration gradient of the electroactive species at the electrode surface. Consequently, the diffusion layer width is:

$$\Delta X = zFADC_{bulk} / i_{peak} \qquad (2)$$

where $z$ is the number of electrons transferred per electroactive species in the redox reaction, $F$, $A$ and $D$ are the Faraday constant, the surface area and the diffusion coefficient, respectively. Knowing that the diffusion coefficient for the system under investigation ($Fe(CN)_6^{3-}/Fe(CN)_6^{4-}$ redox at gold surface) is $5 \times 10^{-5}$ cm$^2$ s$^{-1}$, the values of the inner and outer cutoffs can be estimated in accordance with Eq.(2).

The inner cutoff, which indeed indicates the lower boundary of the fractality scale, can be estimated when the diffusion layer has its smallest thickness. This state corresponding to the experiments performed with fastest rate, thus, the inner cutoff can be calculated by incorporating the peak current recorded at the highest scan rate in Eq.(2). Similarly, the outer cutoff, which reflects the upper boundary of the fractality scale, can be calculated by using the lowest scan rate. The results obtained by this approach are summarized in Table 4. It is obvious that the scale of fractality scale is also dependent on the strength of the plastic deformation induced.

It should be emphasized that the values reported for inner and outer cutoffs of the electrode surfaces are not exactly corresponding to the real fractality scale of the electrode surface. In fact, the inner and outer cutoffs estimated correspond to the smallest and largest scales sensed in our experiments. It should be taken into account that methodology limitation does not allow for an extensive investigation over all length scales. For instance, cyclic voltammetric measurements cannot be performed for such investigation with very low or high scan rates due to the methodology limitation. However, this limitation does not provide problem for our purpose, which is a comparative study. In other words, the difference of the fractality scales of various electrode surfaces under investigation indicates the structural changes induced to the electrode surface during the plastic deformation.

The data reported in Table 4 obviously indicate that the plastic deformation induced to the Pd electrode causes the decrease of the inner cutoff. This suggests that for the Pd electrode subjected to severe plastic deformation, smaller rough structures were formed on the surface. On the other hand, since the outer cutoff is also increased as a result of the plastic deformation (Table 4), it leads to conclusion that the plastic deformation is able to increase the fractality scale. This means that the occurrence of plastic deformation in the metallic system under investigation causes the formation of more complicated surfaces as the rough structure can be sensed in a wider range of the length scales. Of course, it is not possible to monitor structural changes of the surface simply. In fact, it is just indicative of the fact that the fractality scale changes as well as the fractal dimension as a result of the plastic deformation induced. Knowing this issue is of great importance in this context, which provides further opportunities for monitoring the plastic deformation induced.

## 4. Conclusion

For an object subjected to severe plastic deformation, fractal dimension can be used for quantitative analysis of surfaces to inspect the plastic deformation induced. However, due to sensitivity of the structure formed as a result of plastic deformation, it is very important to use an appropriate method for the determination of the fractal dimension, since the interesting structures generated as a result of plastic deformation may be missed. Gold masking is a simple and efficient approach to avoid possible errors, since after Au covering, the object is considered as a conventional electrode. The available electrochemical methods for Au surfaces are the most reliable cases. The present study was typically performed for the plastic deformation induced by phase transformation in the course of hydrogen injection/extraction into/from Pd; however, the method can be generally used for the investigation of plastic deformations leading to surface changes. The novelty of the present work is not due to study of phase transformation-induced plastic deformation and the gold-masking approach for fractal analysis, since both of them have been previously reported in the literature. The aim is to report an appropriate strategy based on available approaches for such investigations. As typically shown, the strategy proposed is desirable to monitor the plastic deformation. In general, the approach proposed is useful for inspecting the solid-state structural changes, since the method is simple and can be used by solid-state physicists (generally non-electrochemists).

**References**

[1]   S. Szpak, C.J. Gabriel, J.J. Smith, R.J. Nowak, J. Electroanal. Chem. 309 (1991) 273.

[2]   F.A. Lewis, The Palladium Hydrogen System, Academic Press, London, 1967.

[3]   H. Peisl, in G. Alefeld and J. Volkl (eds.), Hydrogen in Metals, vol. I, Springer-Verlag, Berlin, 1978, p. 53.

[4]   E. Wicke, Z. Phys. Chem. 143 (1985) 1.

[5]   R.W.K. Honeycombe, The Plastic Deformation of Metals, Edward Arnold, 1984.

[6]   P.R. Roberge, K.R. Trethewey, J. Appl. Electrochem. 25 (1995) 962.

[7]   J.-N. Han, M. Seo, S.-I. Pyun, J. Electroanal. Chem. 514 (2001) 118.

[8]   A. Eftekhari, Phil. Mag. Lett., in press.

[9]   A. Eftekhari, Electrochim. Acta 48 (2003) 2831.

[10]  A. Eftekhari, Appl. Surf. Sci. 220 (2003) 346.

[11]  A. Eftekhari, Colloids Sur. B: Biointerfaces 32 (2003) 375.

[12]  M. H. Pournaghi-Azar, H. Dastangoo, J. Electroanal. Chem. 523 (2002) 26.

[13]  A. Eftekhari, J. Electroanal. Chem., 558 (2003) 75.

[14]  A. Eftekhari, Appl. Surf. Sci., doi: 10.1016/j.apsusc.2003.12.010.

[15]  B.B. Mandelbrot, Fractal Geometry of Nature, Freeman, San Francisco, 1983.

[16]  T. Pajkossy, L. Nyikos, Electrochim. Acta 34 (1989) 171.

[17]  T. Pajkossy, J. Electroanal. Chem. 300 (1991) 1.

[18]  M. Stromme, G.A. Niklasson, C.G. Granqvist, Solid State Commun. 96 (1995) 151.

[19]  M. Stromme, G.A. Niklasson, C.G. Granqvist, Phys. Rev. B 52 (1995) 14192.

[20]  A. Eftekhari, J. Electrochem. Soc., in press.

[21]  A. Czerwinski, I. Kiersztyn, M. Grden, J. Czapla, J. Electroanal. Chem. 471 (1999) 190.

Table 1. Fractal analysis of different electrodes based on gold masking approach.

| Electrodes | Cyclic voltammetry | | Chronoamperometry | |
|---|---|---|---|---|
| | α | $D_f$ | α | $D_f$ |
| Au [a] | 0.509 ± 0.001 | 2.018 | 0.513 ± 0.002 | 2.026 |
| untreated Pd [b] | 0.546 ± 0.001 | 2.092 | 0.559 ± 0.002 | 2.118 |
| $PdH_0$ [c] | 0.608 ± 0.002 | 2.216 | 0.612 ± 0.003 | 2.224 |
| $PdH_{0.23}$ [c] | 0.630 ± 0.002 | 2.260 | 0.638 ± 0.004 | 2.276 |
| $PdH_{0.46}$ [c] | 0.673 ± 0.003 | 2.346 | 0.680 ± 0.004 | 2.360 |
| $PdH_{0.6}$ [c] | 0.732 ± 0.003 | 2.464 | 0.745 ± 0.004 | 2.490 |

[a] Au-coated Au electrode. Au covered by gold-masking approach on a smooth Au surface (with a roughness factor lesser than 2).

[b] A polished Pd electrode, which was not subject of potentiostatic electrochemical treatment.

[c] The notations are related to the pre-charged states. All four electrodes were subject of potentiostatic condition to extract hydrogen.

Table 2. The influence of discharge potential applied in potentiostatic hydrogen extraction on the surface structure of the Pd electrode.

| Electrode | Discharge potential V vs. SCE | Fractal parameter $\alpha$ | Fractal dimension $D_f$ |
|---|---|---|---|
| $PdH_0$ | 0.40 | $0.607 \pm 0.001$ | 2.214 |
|  | 0.60 | $0.608 \pm 0.003$ | 2.217 |
|  | 0.80 | $0.611 \pm 0.002$ | 2.222 |
|  | 0.90 | $0.612 \pm 0.002$ | 2.224 |
| $PdH_{0.6}$ | 0.40 | $0.741 \pm 0.001$ | 2.482 |
|  | 0.60 | $0.734 \pm 0.002$ | 2.468 |
|  | 0.80 | $0.743 \pm 0.002$ | 2.487 |
|  | 0.90 | $0.745 \pm 0.002$ | 2.490 |

* The notations are related to the pre-charged states. Both electrodes were subject of potentiostatic condition to extract hydrogen.

Table 3. The effect of potential scan rate of potentiodynamic hydrogen extraction on the surface structure of the Pd electrode. The discharging process was carried out by sweeping the potential from 0.5 to 1.5 V vs. SCE.

| Electrode | Scan rate / mV s$^{-1}$ | Fractal dimension $D_f$ |
|---|---|---|
| PdH$_0$ | 10 | 2.242 |
|  | 30 | 2.228 |
|  | 50 | 2.221 |
| PdH$_{0.6}$ | 10 | 2.502 |
|  | 30 | 2.494 |
|  | 50 | 2.487 |

* The notations are related to the pre-charged states. Both electrodes were under potentiodynamic condition to extract hydrogen.

Table 4. The inner and outer cutoffs of the Pd electrode subjected to plastic deformation with different strengths.

| Electrodes | Cutoffs indicating the fractality scale | |
|---|---|---|
| | Inner cutoff / μm | Outer cutoff / μm |
| $PdH_0$ | 1.09 | 5.21 |
| $PdH_{0.23}$ | 0.84 | 5.98 |
| $PdH_{0.46}$ | 0.81 | 5.81 |
| $PdH_{0.6}$ | 0.78 | 6.46 |

Figure Caption(s):

Fig. 1. Typical log $I_p$ – log v (upper curve) and log $I$ – log $t$ (curve) plots for the Au-covered Pd electrode (initially pre-charged to $PdH_{0.6}$) obtained from cyclic voltammetric and chronoamperometric measurements, respectively.

Figure 1

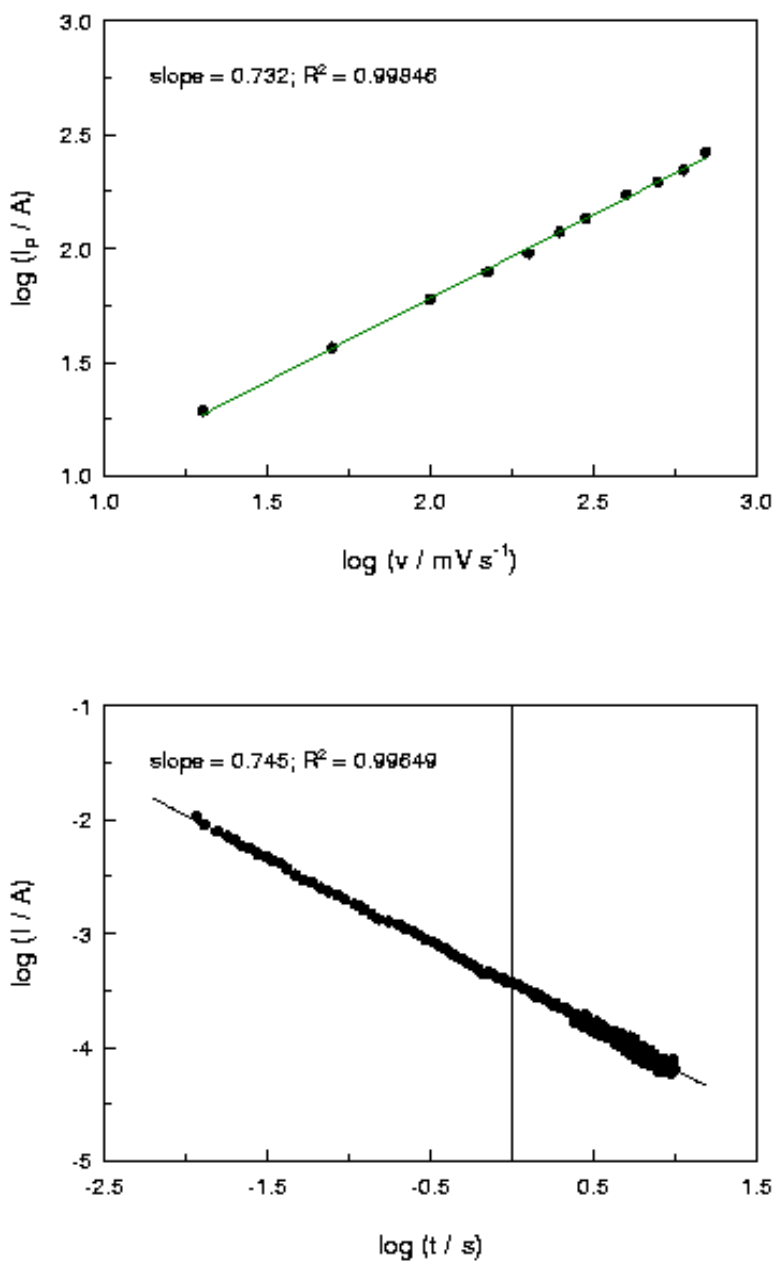